\documentclass[aps,pra,reprint,amsmath,amssymb,superscriptaddress,nobibnotes]{revtex4-2}

\newcommand{\mycomment}[1]{}
\usepackage{natbib}

\renewcommand\bibsection{\section{\bibname}}
\usepackage{graphicx}
\usepackage{color}
\usepackage{physics}
\usepackage{amsmath}
\usepackage{mathtools}
\usepackage{mathtools, nccmath}
\usepackage{array}
\usepackage{cellspace}
\usepackage{makecell}
\usepackage{hyperref}
\usepackage{graphicx,lipsum}% http://ctan.org/pkg/{graphicx,lipsum}
\usepackage{tikz}
\usepackage{lipsum}

\usepackage{expl3}
\ExplSyntaxOn
\int_zero_new:N \g__prg_map_int
\ExplSyntaxOff

\usetikzlibrary{calligraphy}

\begin{document}

%\title{Experimental investigation of entanglement swapping via lossy quantum channels}
\title{Entanglement swapping via lossy channels using photon-number-encoded states}

\author{Wan Zo}
\email{zowan.physics@gmail.com}
\affiliation{Center for Quantum Information, Korea Institute of Science and Technology (KIST), Seoul, 02792, Republic of Korea}
\affiliation{Department of Physics, Yonsei University, Seoul, 03722, Republic of Korea}

\author{Bohdan Bilash}
\affiliation{Center for Quantum Information, Korea Institute of Science and Technology (KIST), Seoul, 02792, Republic of Korea}
\affiliation{Division of Quantum Information, KIST School, Korea University of Science and Technology, Seoul 02792, Republic of Korea}

\author{Donghwa Lee}
\affiliation{Center for Quantum Information, Korea Institute of Science and Technology (KIST), Seoul, 02792, Republic of Korea}

\author{Yosep Kim}
\affiliation{Department of Physics, Korea University, Seoul, 02841, Republic of Korea}

\author{Hyang-Tag Lim}
\affiliation{Center for Quantum Information, Korea Institute of Science and Technology (KIST), Seoul, 02792, Republic of Korea}
\affiliation{Division of Quantum Information, KIST School, Korea University of Science and Technology, Seoul 02792, Republic of Korea}

\author{Kyunghwan Oh}
\affiliation{Department of Physics, Yonsei University, Seoul, 03722, Republic of Korea}

\author{Syed M. Assad}
\affiliation{Centre for Quantum Computation and Communication Technology, Department of Quantum Science, Australian National University, Canberra, ACT, 2601, Australia}
\affiliation{A*STAR Quantum Innovation Centre (Q.InC), Institute of Materials Research and Engineering (IMRE), Agency for Science Technology and Research (A*STAR), 138634, Singapore}

\author{Yong-Su Kim}
\email{yong-su.kim@kist.re.kr}
\affiliation{Center for Quantum Information, Korea Institute of Science and Technology (KIST), Seoul, 02792, Republic of Korea}
\affiliation{Division of Quantum Information, KIST School, Korea University of Science and Technology, Seoul 02792, Republic of Korea}

\date{\today}

\begin{abstract}
\noindent	
Entanglement shared between distant parties is a key resource in quantum networks. However, photon losses in quantum channels significantly reduce the success probability of entanglement sharing, which scales quadratically with the channel transmission. Quantum repeaters using entanglement swapping can mitigate this effect, but usually require high-performance photonic quantum memories to synchronize photonic qubits. In this work, we theoretically and experimentally investigate an entanglement swapping protocol using photon-number-encoded states that can effectively alleviate quantum channel losses without requiring photonic quantum memories. We demonstrate that the protocol exhibits a success probability scaling linearly with the channel transmission. Furthermore, we show that while unbalanced channel losses can degrade the shared entanglement, this effect can be compensated by optimally adjusting the initial entangled states. Our results highlight the potential of photon-number encoding for realizing robust entanglement distribution in lossy quantum networks.

%Sharing entanglement between distant parties is a key resource in quantum networks. Since quantum states cannot be cloned or amplified, photon losses significantly reduce the success probability of entanglement sharing. Entanglement swapping using quantum repeaters can mitigate the effect of photon losses; however, it usually requires high-performance photonic quantum memories to synchronize the arrival time of photonic qubits from distant transmitters. Here, we have theoretically and experimentally investigated an entanglement swapping protocol using {\it photon-number-encoded states} and have demonstrated that quantum channel losses can be effectively alleviated without photonic quantum memories. We have also shown that unbalanced quantum channel losses can degrade the shared entanglement; however, this can be managed by adjusting the initial states.
\end{abstract}

\maketitle

\section{Introduction}

\begin{figure}[t]
	\centering\includegraphics[width=8.6cm]{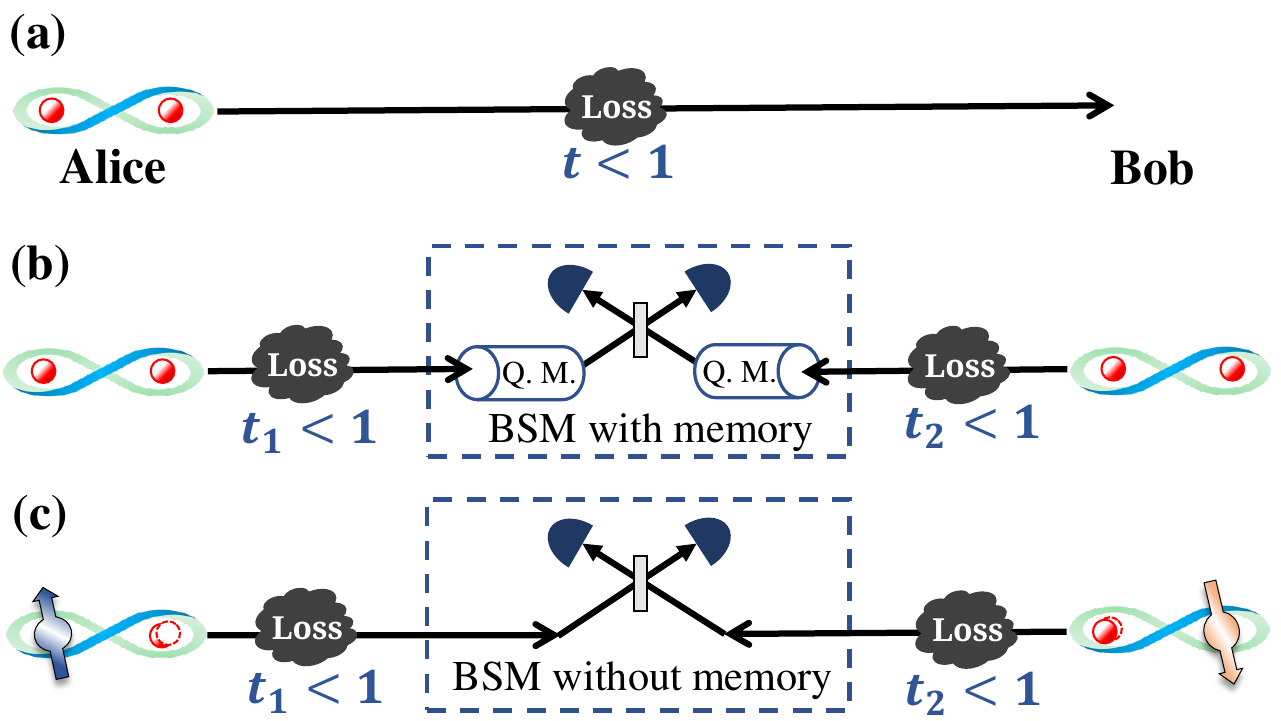}
	\caption{Sharing entanglement between distance parties. $t$ represent transmittivity of total channel. (a) Direct transmission, (b) Conventional entanglement swapping using photonic quantum memories, and (c) Entanglement swapping using photon-number encoded states.}
	\label{fig_idea}
\end{figure}

Entanglement between distant parties is a keystone for many network-based quantum information applications such as quantum communication, networked quantum sensing and quantum computing~\cite{hensen15,gyongyosi21,hong21,pompili21,azuma23,kim24}. Since entanglement cannot be generated by local operations and classical communication, distant parties cannot establish shared entanglement without flying qubits traveling between them. A straightforward way to share entanglement between distant parties would be a party (Alice) generates entangled photons and sends one of the photons to the other party (Bob) via an optical channel, see Fig.~\ref
{fig_idea}(a). The success probability of entanglement sharing scales $O(t^2)$ where $t$ is the amplitude transmittivity of the quantum channel, and it will be assumed to be real and $0\le t\le1$ without loss of generality. This scaling and the upper limit is known as a fundamental limit of quantum communications without quantum repeaters and known as the Pirandola-Laurenza-Ottaviani-Banchi (PLOB) bound~\cite{plob}.

The PLOB bound can be surpassed by using quantum repeaters~\cite{azuma23}. As shown in Fig.~\ref{fig_idea}(b), in a quantum-repeater-based entanglement swapping protocol, Alice and Bob generate two-photon entanglement and send one of photons to the central node via optical channels that have channel transmittivities $t_1$ and $t_2$ where $t_1 t_2=t$. When the central node successfully performs entanglement measurement, Alice and Bob can share entanglement. Note that the entanglement measurement usually requires simultaneous arrival of two single-photons from Alice and Bob~\cite{kim18}. If the central node has ideal photonic quantum memories which store and retrieve a photonic qubit at a desirable timing, it can store the early arriving photon until the later one arrives and performs entanglement measurement. Therefore, the success probability of entanglement sharing is given as $O(t)$. If no such ideal photonic quantum memories are available, however, the entanglement measurement can be performed only when two photons from Alice and Bob simultaneously arrived at the central node. In this case, the successful entanglement sharing probability is still $O(t^2)$. Therefore, high quality photonic quantum memories are essential for the quantum repeater assisted entanglement sharing scenario. Despite remarkable recent development, practical photonic quantum memories for quantum repeater are yet beyond present technologies~\cite{qmem1, qmem2,qmem3,qmem4}.

%%%%%%%%%%%%%%%%%%%%%%%%%%%%
%%%%%%%%%%%%%%%%%%%%%%%%%%%%

In quantum key distribution (QKD), the idea of entanglement swapping is applied to measurement-device-independent QKD (MDI-QKD) where successful entanglement measurement in a central node establishes the secret keys between distant parties~\cite{mdiqkd2,mdiqkd,pnp-mdi,park18, lee20}. Without photonic quantum memories, the secret key rate scales $O(t^2)$ since both photons from Alice and Bob should simultaneously arrive at the central node. The quantum memory requirement, however, has been relaxed by the novel idea of Twin-Field QKD (TF-QKD)~\cite{tfqkd}. TF-QKD is a type of MDI-QKD where the successful entanglement measurement happens even when only one of Alice's or Bob's photons arrive at the central node. With this remarkable property, the secret key rate of TF-QKD is given as $O(t)$, and thus, can surpass the PLOB bound. TF-QKD has been implemented over long distances that are intractable via conventional QKD protocols~\cite{liu19,chen20,liu21,chen21,wang22,liu23}.

The essence of TF-QKD is that the bit information is encoded into the {\it number of photons}, i.e., vacuum or non-vacuum states, and thus the entanglement measurement can be implemented with only one of photons from either Alice or Bob~\cite{ozlem22}. The {\it photon-number encoding} can be applied to more general quantum network scenarios such as entanglement swapping. For instance, Alice and Bob prepare spin-photon entanglement where the photonic qubit is encoded into the photon-number state and utilized for entanglement swapping to establish entanglement between distant spin qubits, see Fig.~\ref{fig_idea}(c). Such protocol has been first proposed by Cabrillo {\it et al}~\cite{cabrillo99}, and widely implemented in various physical systems to share entanglement between distant quantum nodes~\cite{dlcz,yang16,kutluer17,laplane17,humphreys18,hermans23}. Despite the significant interest in the photon-number-encoded entanglement swapping protocol, its performance and robustness against optical channel losses have not been thoroughly investigated.

In this work, we theoretically and experimentally investigate the photon-number-encoding in the context of entanglement swapping via lossy optical channels. In particular, we analyze the quality of entanglement and the success probability. We also find that by optimizing the initial states and compensating the channel loss imbalance, we can achieve improved entanglement distribution and overcome inherent channel losses. We also report the proof-of-principle experiment to verify our theoretical findings.

\section{Theory}

\begin{figure}[t]
	\centering\includegraphics[width=8.6cm]{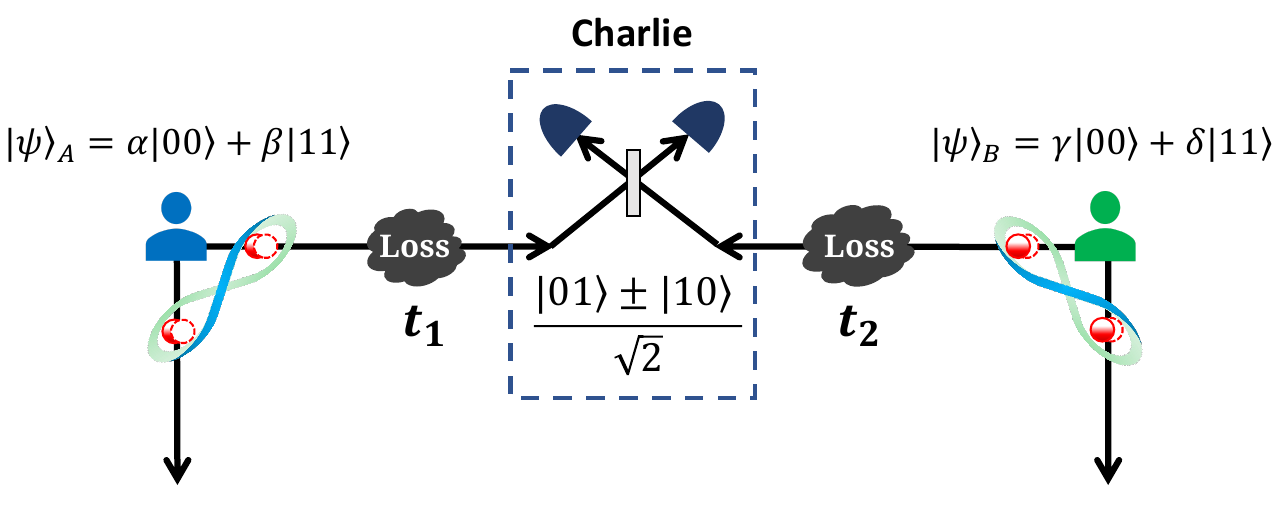}
	\caption{Schematic diagram of the entanglement swapping protocol using photon-number-encoded states.}
	\label{fig_scheme}
\end{figure}

Figure~\ref{fig_scheme} shows the outline of the entanglement swapping protocol using photon-number encoded states via lossy channels. The channel transmittivities are $t_1$ (between Alice and Charlie) and $t_2$ (between Bob and Charlie), respectively. Alice and Bob prepare initial entanglement as
\begin{eqnarray}
	|\psi\rangle_A&=&\alpha|00\rangle+\beta|11\rangle,\nonumber\\
	|\psi\rangle_B&=&\gamma|00\rangle+\delta|11\rangle, 
	\label{input}
\end{eqnarray}
where $|\alpha|^2+|\beta|^2=|\gamma|^2+|\delta|^2=1$. Note that one of the qubits is encoded in the photon-number while the other can be encoded any physical states, e.g., spin states. Now, Alice and Bob transmit the {\it photon-number-encoded} photonic qubits to the third-party, Charlie, via lossy channels. Since the lossy channel corresponds to the amplitude damping channel in photon-number encoding, it can be described as~\cite{lee11,kim12}
\begin{eqnarray}
|\psi\rangle_A |0\rangle_{E_1} &\rightarrow&|\psi\rangle_{AC_1E_1} \nonumber\\
&=& \alpha|0\rangle_A |0\rangle_{C_1} |0\rangle_{E_1} 
+\beta r_1|1\rangle_A |0\rangle_{C_1} |1\rangle_{E_1} \nonumber\\
&&+\beta t_1|1\rangle_A |1\rangle_{C_1} |0\rangle_{E_1},\\
|0\rangle_{E_2} |\psi\rangle_{B} &\rightarrow& |\psi\rangle_{E_2C_2 B} \nonumber\\
&=& \gamma|0\rangle_{E_2} |0\rangle_{C_2} |0\rangle_B 
+ \delta r_2|1\rangle_{E_2} |0\rangle_{C_2} |1\rangle_B\nonumber\\
&&+\delta t_2|0\rangle_{E_2} |1\rangle_{C_2} |1\rangle_B.	\nonumber
\label{lossy transform}
\end{eqnarray}
where $t_{1}^2+r_{1}^2=1$ and $t_{2}^2+r_{2}^2=1$. The subscripts denote Alice ($A$), Bob ($B$), Charlie ($C_1$ and  $C_2$), and environment (E). 

By tracing out the environment $E$, one can obtain the effective states shared by Alice (Bob) and Charlie, $\rho_{AC_1}~(\rho_{C_2B})$. The overall state shared by Alice, Bob and Charlie is $\rho_{ABC}=\rho_{A C_1}\otimes\rho_{C_2 B}$. Now, Charlie performs the Bell state measurement (BSM) onto $|\Psi^{\pm}\rangle=\frac{1}{\sqrt{2}}\left(|01\rangle \pm|10\rangle\right)$, and the state between Alice and Bob after the successful BSM becomes
%  %  %  %  %  %  %  %  %  %  %
\begin{equation}
\rho_{AB}\sim Tr_{C_1C_2}[\Pi_{BSM}\rho_{ABC}\Pi_{BSM}^\dagger],
\label{eq_BSM}
\end{equation}
%  %  %  %  %  %  %  %  %  %  %
where $\Pi_{BSM}= I_A\otimes |\Psi^{\pm}\rangle_{C_1C_2}\langle \Psi^{\pm}|_{C_1C_2}\otimes I_B$. Note that the BSM on photon-number-encoded photonic qubits can be easily implemented by one and only one single-photon detection either at $C_1$ or $C_2$ after a beamsplitter (BS). With the success of BSM, the remaining quantum state shared by Alice and Bob is given as
%  %  %  %  %  %  %  %  %  %  %
\begin{equation}
	\rho^\pm_{AB} 
	= \frac{1}{\mathcal{N}}\begin{pmatrix}
		0& 0& 0&0 \\
		0& |\alpha\delta|^2 t_2^2& \pm \alpha\beta^*\gamma^*\delta t_1t_2& 0\\
		0& \pm \alpha^*\beta\gamma\delta^* t_1t_2& |\beta\gamma|^2 t_1^2& 0\\
		0& 0& 0&\rho_{44}
	\end{pmatrix},
	\label{final matrix}	
\end{equation}
%  %  %  %  %  %  %  %  %  %  %
where $\rho_{44}=|\beta\delta|^2\{t_1^2(1-t_2^2)+t_2^2(1-t_1^2)\}$ and $\mathcal{N}=|\alpha\delta|^2 t_2^2+|\beta\gamma|^2 t_1^2+\rho_{44}$ is the normalization constant.

\begin{figure}[b]
	\centering\includegraphics[width=8.6cm]{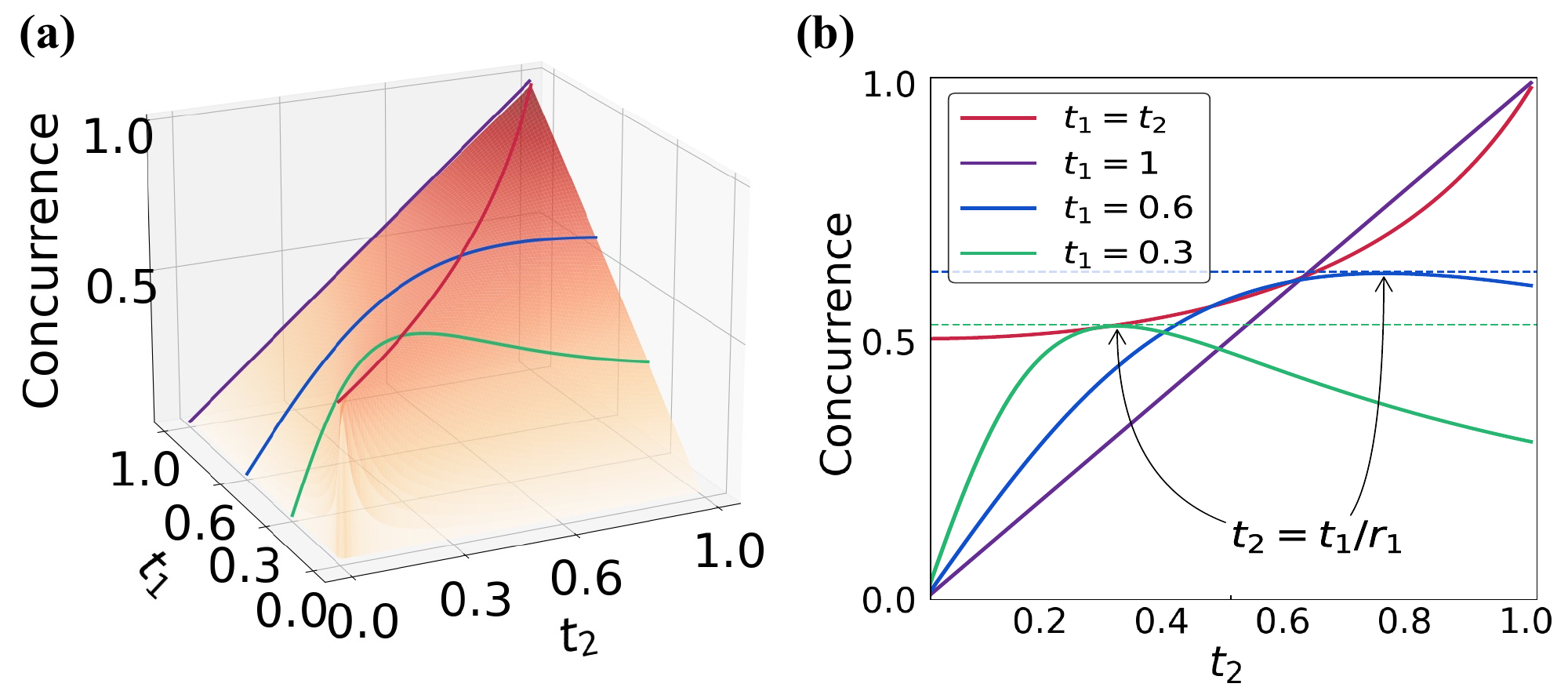}
	\caption{Concurrence of $\rho_{AB}$ in terms of channel transmittivities $t_1$ and $t_2$ when the initially prepared states by Alice and Bob are maximally entangled state. For $t_1<r_1$, concurrence becomes maximum at $t_2=t_1/r_1$.}
	\label{fig_concur1}
\end{figure}

The amount of entanglement for two-qubit states $\rho_{AB}$ can be quantified by concurrence and it is calculated as~\cite{concur}:
%  %  %  %  %  %  %  %  %  %  %
\begin{equation}
	C = \frac{2|\alpha\beta\gamma\delta t_1 t_2|}{\mathcal{N}}.
	%C = 2\mathcal{N}|\alpha\beta\gamma\delta t_1 t_2|.
	\label{eq_con}
\end{equation}
%  %  %  %  %  %  %  %  %  %  %
Figure~\ref{fig_concur1}(a) shows the concurrence of $\rho_{AB}$ in terms of the channel transmittivities $t_1$ and $t_2$ with the maximally entangled initial states $|\psi\rangle_A$ and $|\psi_B\rangle$, i.e., $|\alpha|=|\beta|=|\gamma|=|\delta|=1/\sqrt{2}$. It shows that concurrence generally decreases as the channel transmittivities decrease or equivalently the channel losses increase. This can be understood as an increasing effect of $\rho_{44}$ term with decreasing channel transmittivities. Indeed, $\rho_{44}$ term corresponds the case when both Alice and Bob have sent single photons, but Charlie received only a single-photon due to the optical loss, and its contribution increases as the loss increases.

Another important aspect to the resultant entanglement is balancing the channel losses. In Fig.~\ref{fig_concur1}(a), we present different scenarios of $t_1$ and $t_2$ with different colored curves. For clarity, we present the same curves with respect to $t_2$ in 2D plots, see Fig.~\ref{fig_concur1}(b). It is noteworthy that concurrence does not vanish even when both $t_1$ and $t_2$ asymptotically approaches to 0 ($t_1,t_2\rightarrow0$) as long as they are balanced, i.e., $t_1=t_2$. On the other hand, when the channel transmittivities are different, ($t_1\neq t_2$), vanishing one of the transmittivities ($t_1~{\rm or}~t_2\rightarrow0$) causes zero concurrence. Moreover, it sometimes happens that larger optical loss can result in larger amount of entanglement. For instance, when $t_1=0.3$, $t_2=0.3$ results larger concurrence than $t_2=0.6$ or even $t_2=1$. In specific, concurrence is maximized at $t_2=t_1/r_1$ for $t_1<r_1$. This indicates that balancing the channel transmittivities plays an important role in entanglement swapping protocol using photon-number encoded states.

%%%%%%%%%%%%%%%%%%%%%%%%%%%%
%%%%%%%%%%%%%%%%%%%%%%%%%%%%

Figure~\ref{fig_concur1} shows that, a maximally entangled state, i.e., $C=1$, cannot be shared if Alice and Bob begin the entanglement swapping protocol with maximally entangled states, i.e., $|\alpha|=|\beta|=|\gamma|=|\delta|=\frac{1}{\sqrt{2}}$, unless the quantum channels do not have loss, i.e., $t_1=t_2=1$. This is because of the non-vanishing $\rho_{44}$ term in Eq.~\eqref{final matrix}. It is remarkable that $\rho_{44}$ term becomes negligible comparing to other terms when $|\alpha| \gg |\beta|$ and $|\gamma| \gg |\delta|$. Therefore, in this limit, the final state can be rewritten as a (unnormalized) pure state as
%  %  %  %  %  %  %  %  %  %  %  %  %
\begin{eqnarray}
	\rho^\pm_{AB}
	\xrightarrow[|\alpha|,|\gamma| \gg |\beta|,|\delta|]{}|\psi^{\pm} \rangle_{AB}\sim\alpha \delta t_2 |01\rangle \pm  \beta \gamma t_1 |10\rangle.
	\label{pureform}
\end{eqnarray}
%  %  %  %  %  %  %  %  %  %  %  %  %
It is notable that regardless of $t_1$ and $t_2$, one can always make Eq.~\eqref{pureform} to a maximally entangled state by adjusting the coefficients $\alpha, \beta, \gamma$ and $\delta$ of the initial states. In particular, Eq.~\eqref{pureform} becomes a maximally entangled state when $|\alpha\delta t_2|=|\beta\gamma t_1|$. These conditions can be satisfied as follows:
%  %  %  %  %  %  %  %  %  %  %  %  %
\begin{eqnarray}
	\begin{aligned}
	|\alpha\delta|=\sqrt{\frac{t_1^2}{t_1^2+t_2^2}},~~~|\beta\gamma|=\sqrt{\frac{t_2^2}{t_1^2+t_2^2}}.
		\label{eq_restore}
	\end{aligned}
\end{eqnarray}
%  %  %  %  %  %  %  %  %  %  %  %  %
Now let us investigate how the success probability of sharing entanglement scales with the channel loss. For simplicity, let us consider the case where $t_1=t_2=\sqrt{t}$. It represent the case where Charlie is at the middle of Alice and Bob and the entire channel transmittivity is $t$. Then, Eq.~\eqref{pureform} can be rewritten as
%  %  %  %  %  %  %  %  %  %  %  %  %
\begin{eqnarray}
	|\psi^{\pm} \rangle_{AB}\rightarrow\sqrt{t}\left(\alpha \delta |01\rangle \pm  \beta \gamma |10\rangle\right).
	\label{pureform_loss}
\end{eqnarray}
%  %  %  %  %  %  %  %  %  %  %  %  %
Therefore, the success probability to share entanglement between Alice and Bob scales $O(t)$.

More generally, in the case where two channels have different transmittivities $t_1\neq t_2$, we can express shared quantum state satisfying Eq. \eqref{eq_restore} as follows:
\begin{eqnarray}
	|\psi^{\pm} \rangle_{AB}\rightarrow \sqrt{\frac{2t_1^2 t_2^2}{t_1^2 + t_2^2}}\left(\frac{|01\rangle\pm|10\rangle}{\sqrt{2}}\right).
	\label{eq_unbalanced_state}
\end{eqnarray}
Consequently, the success probability $P_{scc}$ to share maximally entangled state is
\begin{flalign}
	\begin{aligned}
		P_{scc} \propto \frac{2t_1^2 t_2^2}{t_1^2 +t_2^2} \propto t.
		\label{eq_scc_imbalance}
	\end{aligned}
\end{flalign}

\section{Experiment}

%%%%%%%%%%%%%%%%%%%%%%%%%%%
\begin{figure*}[t]
	\centering\includegraphics[width=13cm]{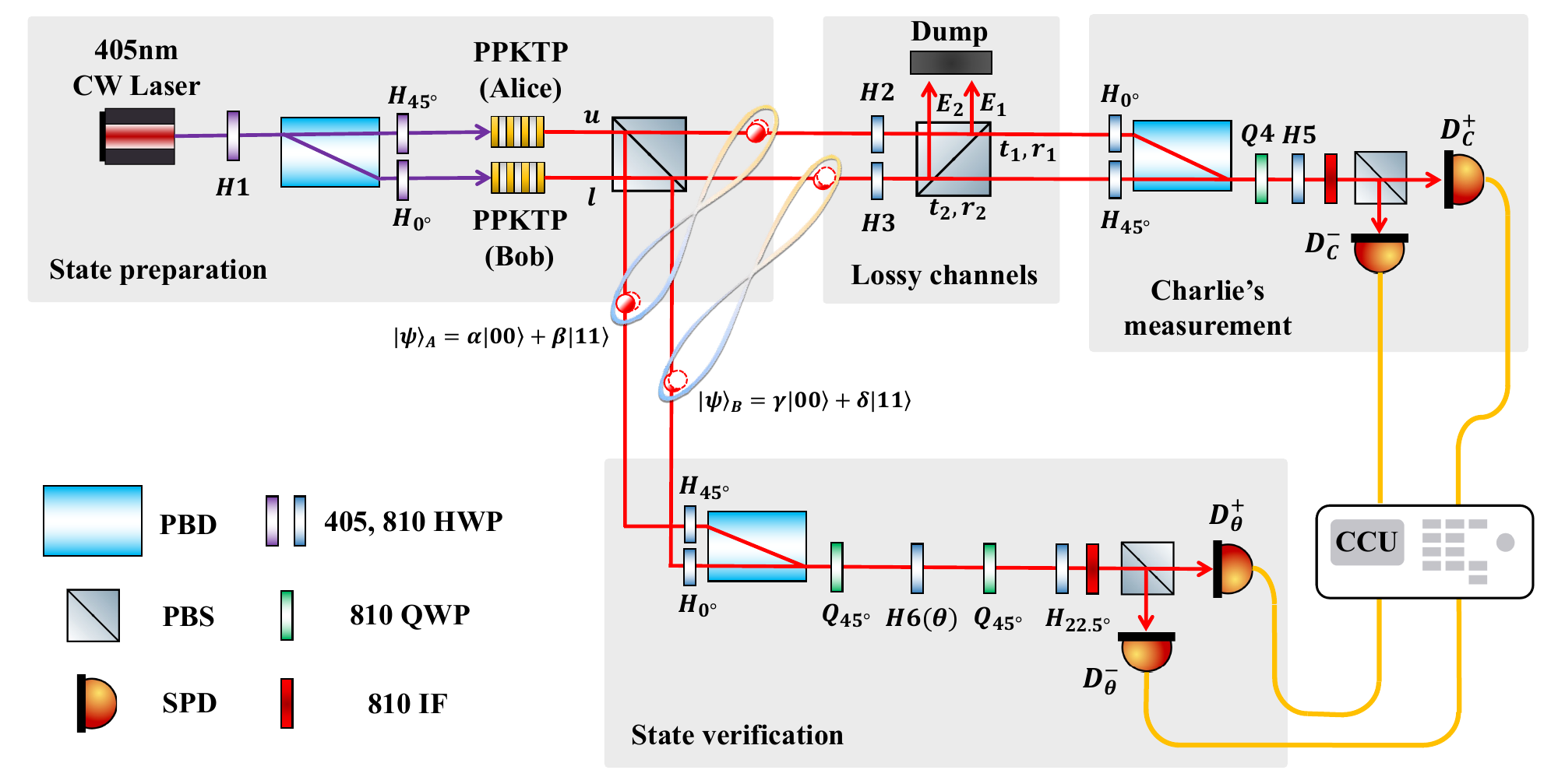}
	\caption{The experiment setup. PBD: polarization beam displacer, PBS: polarizing beamsplitter, SPD: single-photon detector, HWP: half-waveplate, QWP: quarter-waveplate, IF: interference filter, PPKTP: periodically-poled KTP crystal.}
	\label{fig_setup}
\end{figure*}
%%%%%%%%%%%%%%%%%%%%%%%%%%%
Figure~\ref{fig_setup} presents our experiment setup to investigate the entanglement swapping protocol using photon-number encoded states via lossy channels. The continuous-wave (CW) pump laser beam at the wavelength of 405~nm is divided into two spatial modes via a polarizing beam displacer (PBD) which splits and laterally displaces beams according to the polarization. The splitting ratio between two spatial modes is controlled by a half-waveplate (HWP, H1) before the PBD. Then, the two laser beams are used to pump spontaneous parametric down-converion (SPDC) at 10~mm long type-II periodically-polled KTP (PPKTP) crystals. In order to avoid multiple photon-pair generation, we used low pump power of 0.5~mW at the crystals. The SPDC at a lower power regime can be presented as
% % % % % % % % % % % % % % % % % % % % % %
 \begin{equation}
|\psi\rangle_{\rm SPDC}\sim|0\rangle_H|0\rangle_V+\xi|1\rangle_H|1\rangle_V,
\label{spdc}
\end{equation}
% % % % % % % % % % % % % % % % % % % % % %
where $|0\rangle$ and $|1\rangle$ are the vacuum and single-photon states and the subscripts $H$ and $V$ denote the horizontal and vertical polarization states, respectively. Therefore, by considering each PPKTP crystal belongs to each party, Alice and Bob, the SPDC process corresponds to the initial entangled state generation of Eq.~\eqref{input}. Note also that since $|\xi|\ll1$ for a low pump power, Eq.~\eqref{spdc} satisfies the condition of $|\alpha|\gg|\beta|$ and $|\gamma|\gg |\delta|$.

Then, the SPDC photon pairs are divided by a polarizing beam splitter (PBS) and the transmitted photons are sent to Charlie who performs Bell state measurement via lossy quantum channels while the reflected photons are kept by each party. The quantum channel loss is demonstrated by HWPs (H2 and H3) and a PBS. The transmittivities $t_1$, $t_2$ and reflectivities corresponding to photon losses $r_1$, $r_2$ are represented in the Fig.~\ref{fig_setup}.

The Bell state measurement apparatus performs the projective measurement onto $|\Psi^\pm\rangle=\frac{1}{\sqrt{2}}\left(|0\rangle_u|1\rangle_l\pm|1\rangle_u|0\rangle_l\right)$ where the subscripts $u$ and $l$ denote the upper and lower spatial modes, respectively. To this end, the polarization state at the lower path is converted to vertical using a HWP and two spatial modes are combined into a single spatial mode using a PBD. Note that it corresponds to a spatial mode to polarization mode converter, so the Bell state measurement would be performed onto 
$|\Psi^\pm\rangle=\frac{1}{\sqrt{2}}\left(|0\rangle_u|1\rangle_l\pm|1\rangle_u|0\rangle_l\right)$. Now various two-qubit projective measurements can be performed by waveplates followed by a PBS and single-photon detectors at the outputs. Here, we performed two entanglement measurements onto $|\Psi^\pm_X\rangle=\frac{1}{\sqrt{2}}(|01\rangle\pm|10\rangle)$ and $|\Psi^\pm_Y\rangle=\frac{1}{\sqrt{2}}(|01\rangle\pm i|10\rangle)$ and a separable measurement onto $|\Psi^+_Z\rangle=|01\rangle$ and $|\Psi^-_Z\rangle=|10\rangle$.

Upon with the Bell state measurement outcomes of $|\Psi^\pm_X\rangle$ and $|\Psi^\pm_Y\rangle$ at Charlie, Alice and Bob share entangled states. Since logical $|0\rangle$ and $|1\rangle$ are encoded in the photon-number states at Alice and Bob, the entanglement verification requires joint measurement similar to that of Bell state measurement apparatus. In particular, we performed projective measurements onto 
% % % % % % % % % % % % % % % % % % % % % %
 \begin{equation}
|\Psi^{\pm}(\theta)\rangle=\frac{1}{\sqrt{2}}\left(|01\rangle\pm e^{i\theta}|10\rangle\right).
\label{projection}
\end{equation}
% % % % % % % % % % % % % % % % % % % % % %
The phase factor $e^{i\theta}$ is implemented using combination of waveplates where a HWP with $\theta/4$ (H6($\theta$) in Fig.~\ref{fig_setup}) between two quarter waveplates at 45$^\circ$~\cite{yoon19}. 

Single-photon detectors $D^\pm_{C}$ and $D^\pm_\theta$ correspond to projection measurement $|\Psi^{\pm}_{X,Y,Z}\rangle\langle \Psi^{\pm}_{X,Y,Z}|$ at Charlie's entanglement measurement and $|\Psi^{\pm}(\theta)\rangle\langle \Psi^{\pm}(\theta)|$ at the state verification, respectively. Thus, the coincidence detections between $D^\pm_{C}$ and $D^\pm_\theta$ reveal the properties of the shared quantum state between Alice and Bob depending the Charlie's entanglement measurement results. The single and coincidence counts are registered by a home-made coincidence count unit~\cite{park15,park21}.

\begin{figure}[t]
	\centering
	\includegraphics[width=8.6cm]{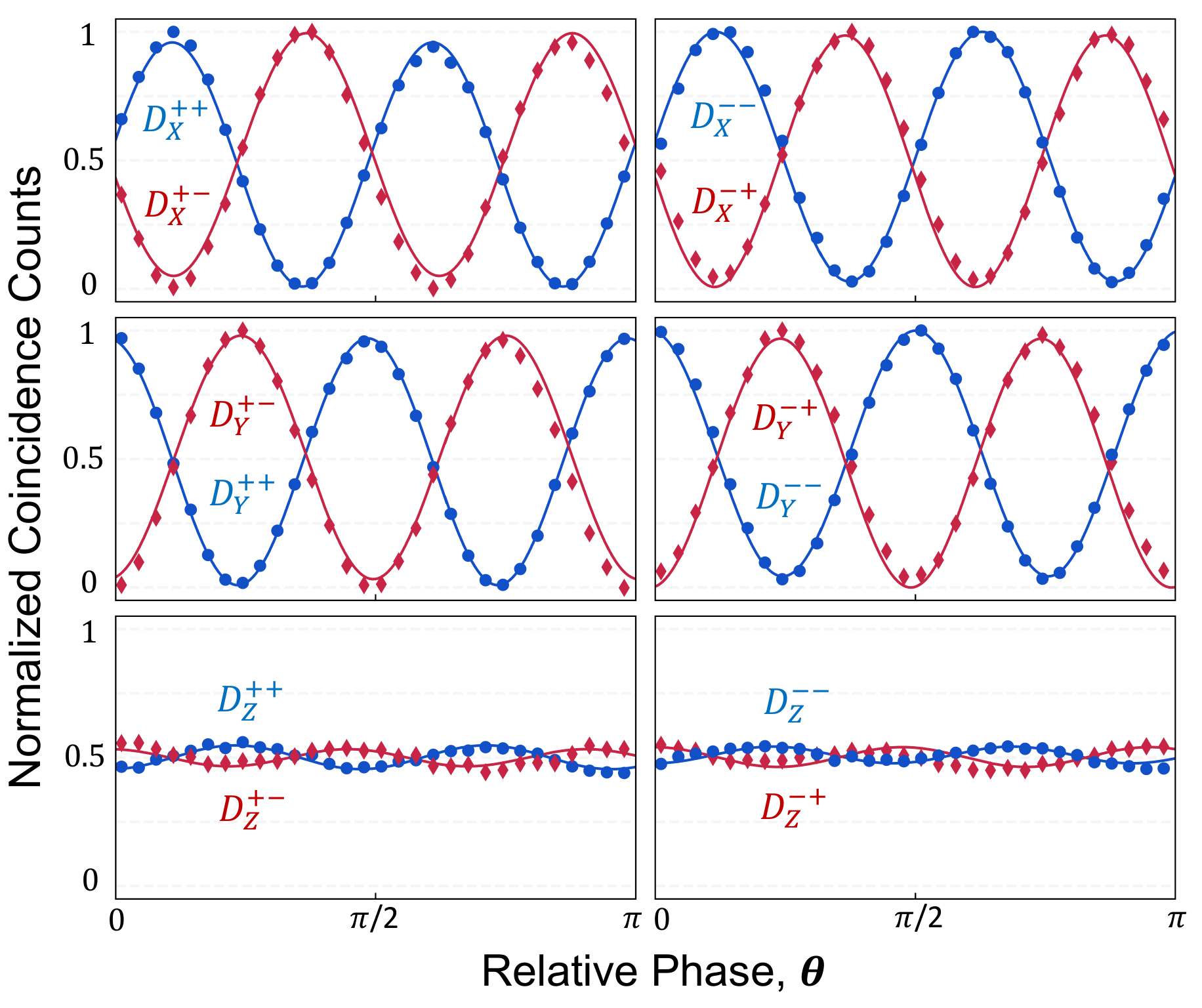}
	\caption{Coincidence counts for the Charlie's measurements $|\Psi^{\pm}_{X}\rangle\langle \Psi^{\pm}_{X}|$, $|\Psi^{\pm}_{Y}\rangle\langle \Psi^{\pm}_{Y}|$ and $|\Psi^{\pm}_{Z}\rangle\langle \Psi^{\pm}_{Z}|$, from top to bottom. When Charlie performs entanglement measurements, the top and middle rows, the coincidence counts reveal clear interference. On the other hand, when Charlie performs separable measurement, interference vanishes (the bottom row).}
	\label{fig_results1}
\end{figure}

The four combinations of coincidence detections $D^{\pm\pm}$ with respect to the relative phase $\theta$ are presented in Fig.~\ref{fig_results1}. Here, for example, $D^{+-}_M$ denotes the coincidence detection between $D_C^+$ and $D_\theta^-$ when the Chalie's measurement is on $|\Psi^{\pm}_M\rangle$ where $M\in\{X,Y,Z\}$. As shown the first and second rows, when Charlie performs entanglement measurement, $|\Psi^{\pm}_{X}\rangle\langle \Psi^{\pm}_{X}|$ and $|\Psi^{\pm}_{Y}\rangle\langle \Psi^{\pm}_{Y}|$, the coincidence detections reveal clear interference. The average interference visibilities for $|\Psi^{\pm}_{X}\rangle\langle \Psi^{\pm}_{X}|$ and $|\Psi^{\pm}_{Y}\rangle\langle \Psi^{\pm}_{Y}|$ projection measurements are $V^+_X=0.962\pm 0.016$, $V^-_X=0.944\pm 0.015$, $V^+_Y=0.949\pm 0.010$ and $V^-_Y=0.968\pm 0.011$, respectively. On the other hand, when Charlie performs separable measurements, $|\Psi^{\pm}_{Z}\rangle\langle \Psi^{\pm}_{Z}|$, the interference fringes vanish. In our experiment, we observe little interference with the visibility of $V_Z=0.075\pm0.010$ and this might be caused by optical axes missalignment. Note that the high two-photon interference visibilities of Fig.~\ref{fig_results1} indicate coherence between $|01\rangle$ and $|10\rangle$ terms and entanglement between them~\cite{vis_intensity}. The expected fidelity with Bell states are estimated as $F_X^+=0.981\pm0.008$, $F_X^-=0.972\pm 0.007$, $F_Y^+=0.974\pm 0.005$ and $F_Y^-=0.984\pm 0.006$, and thus, present the successful entanglement distribution between Alice and Bob via entanglement measurement at Charlie. 

%%%%%%%%%%%%%%%%%%%%%%%%%%%
%%%%%%%%%%%%%%%%%%%%%%%%%%%

%%%%%%%%%%%%%%%%%%%%%%%%%%%
\begin{figure}[t]
\centering
	\includegraphics[width=8.6cm]{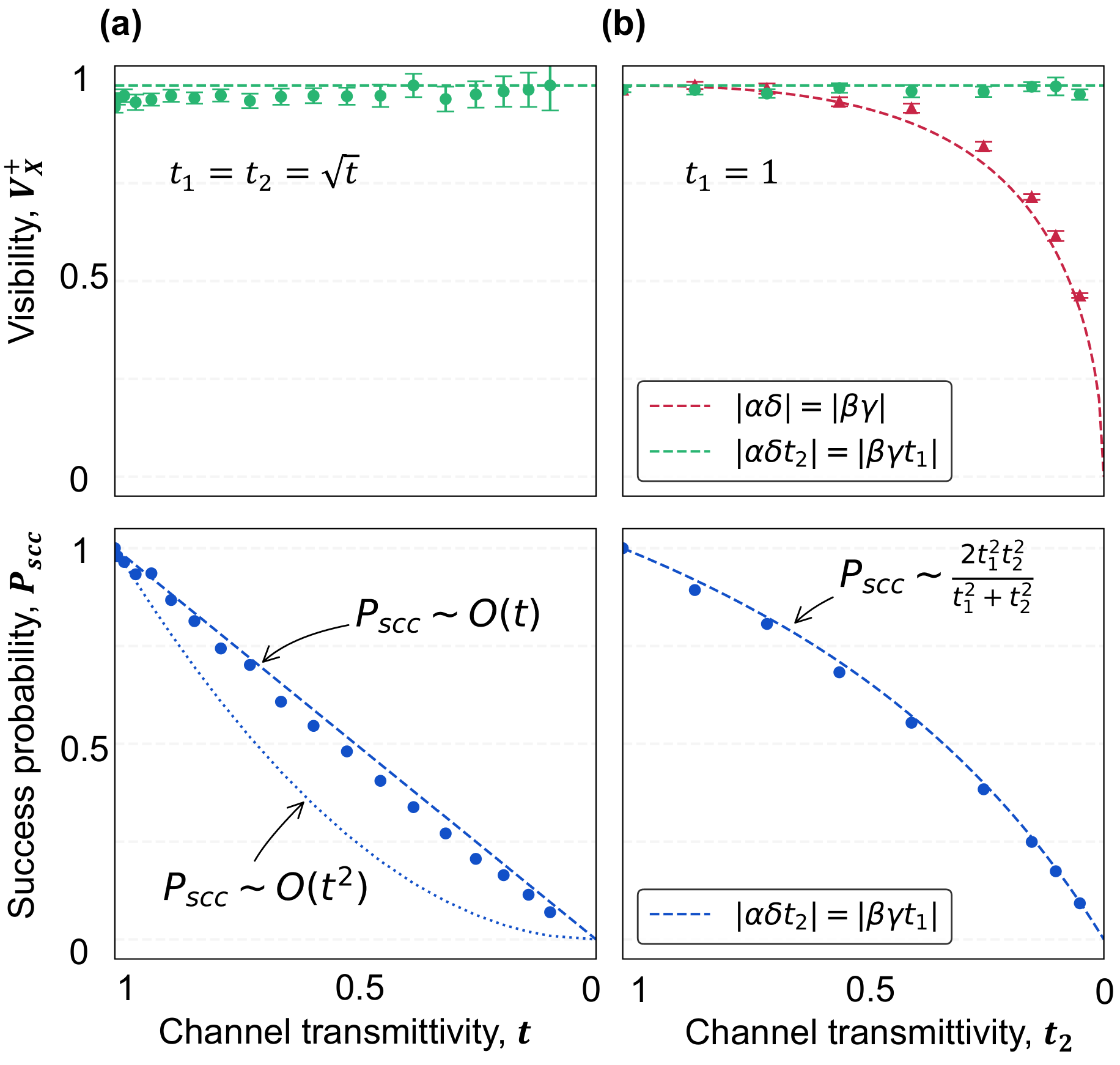}
	\caption{Visibility (top) and the success probability (bottom) with respect to the channel transmittivity. (a) When two channel losses are equal ($t_1=t_2=\sqrt{t}$), and (b) two channel losses are different, i.e., $t_1$ is lossless meaning that $t_1=1$. The success probability was obtained by normalizing the coincidence counts for lossless channels, i.e., $t_1=t_2=1$.}
	\label{fig_results2}
\end{figure}
%%%%%%%%%%%%%%%%%%%%%%%%%%%

We have also experimentally investigated the effect of the channel losses. The amount of loss was experimentally calibrated by monitoring the counts at Chalie's detector while rotating the HWPs for losses (H2 and H3). Figure~\ref{fig_results2} (a) shows the visibility of $V_X^+$ and the success probability with respect to the balanced channel losses, $t_1=t_2$. The visibility is remained over $V_X^+>0.95$ despite the increasing channel losses. The experimental success probability is obtained by normalizing coincidence counts with those with lossless transmission, i.e., $t_1=t_2=1$. Dotted and dashed lines illustrate the theoretical success probability of $O(t^2)$ and $O(t)$, respectively. The experimental results clearly exhibit linear scaling of $O(t)$ so a quadratic improvement compared to the direct transmission.

Figure~\ref{fig_results2}(b) shows the experimentally obtained visibility $V_X^+$ and the success probability when two quantum channels have different losses. In this experiment, we fixed at $t_1=1$ and varied $t_2$. When the two initial states are the same, $|\alpha\delta|=|\beta\gamma|$, the visibility $V_X^+$ decreases as the channel loss imbalance increases, see the red markers and lines of the top graph in Fig.~\ref{fig_results2}(b). This degradation of visibility can be restored by changing the initial states so as to satisfy $|\alpha\delta t_2|=|\beta\gamma t_1|$. In the experiment, such condition was satisfied by changing the pump power ratio of two SPDC crystals with HWP (H1). Note that these states satisfy the condition of Eq.~\eqref{eq_restore}. The restoring visibility is presented as green markers and lines. As we predicted in Eq.~\eqref{eq_scc_imbalance}, however, this increasing quality of entanglement comes with the expense of decreasing the success probability as shown in the bottom graph of Fig.~\ref{fig_results2}(b).

\section{Conclusion}

We have theoretically and experimentally investigated an entanglement swapping protocol utilizing photon-number-encoded states to overcome the detrimental effects of lossy optical channels. Our quantitative analysis demonstrates that the entanglement distribution probability of this protocol scales linearly with the channel transmittivity, providing a significant improvement over direct transmission without requiring quantum memories. Moreover, we have identified that imbalanced channel losses can lead to degradation of the shared entanglement quality. We have also shown that this degradation can be mitigated by judiciously tailoring the initial entangled states. The theoretical predictions are verified by proof-of-principle experiments. Our findings establish photon-number encoding as a promising approach for enabling high-quality entanglement distribution in practical quantum networks via lossy optical channels. 

\section*{Acknowledgements}
\noindent This research was funded by Korea Institute of Science and Technology (2E32941, 2E32971), National Research Foundation of Korea (2023M3K5A1094805, 2022M3K4A1094774), IITP-MSIT (RS-2024-00396999) and KREONET Advanced Research Program Grant from KISTI.

% ORP 2E32941 (2024)
%KIST project (Han) 2E32971 (2024)
% NRF-Quantum simulator 2023M3K5A1094805
% NRF-quantum error correction [International] 2022M3K4A1094774
% IITP-advanced QKD  RS-2024-00396999

\noindent\vspace{0.5\baselineskip}%
\tikz \calligraphy [copperplate] (0,0) -- ++(0.5\textwidth,0) [this stroke style={light,taper=both}];

\renewcommand{\bibsection}{}

%}

\end{document}